# Geometrical Interpretation of Shannon's Entropy Based on the Born Rule

Marko V. Jankovic

*Abstract* – In this paper we will analyze discrete probability distributions in which probabilities of particular outcomes of some experiment (microstates) can be represented by the ratio of natural numbers (in other words, probabilities are represented by digital numbers of finite representation length). We will introduce several results that are based on recently proposed JoyStick Probability Selector, which represents a geometrical interpretation of the probability based on the Born rule. The terms of generic space and generic dimension of the discrete distribution, as well as, effective dimension are going to be introduced. It will be shown how this simple geometric representation can lead to an optimal code length coding of the sequence of signals. Then, we will give a new, geometrical, interpretation of the Shannon entropy of the discrete distribution. We will suggest that the Shannon entropy represents the logarithm of the effective dimension of the distribution. Proposed geometrical interpretation of the

Shannon entropy can be used to prove some information inequalities in an elementary way.

## I. INTRODUCTION

This paper has the main goal to give a simple, intuitive, geometrical interpretation of discrete probability distribution, and based on it, simple, geometrical interpretation of the Shannon entropy. Here, we will analyze discrete probability distributions in which probabilities of particular outcomes of some experiment (microstates) can be represented by the ratio of natural numbers. In other words, probabilities are represented by digital numbers of finite representation length, which is the case in all situations of practical interest. These kinds of probability distributions are used every time we use personal computers to solve some kind of machine learning problem.

We will use this simple geometrical interpretation to give a new geometrical interpretation of the Shannon entropy in order to understand essence of entropy as a measure. In [3] a universal geometric approach to entropy has been purposed. In that paper it was said that entropy could be understood as an ensemble volume. In that case it is difficult to understand how it is possible to compare volumes of different dimensions. Here, we will propose a different approach by showing that the Shannon entropy represents logarithm of the ratio of two combinatorial volumes divided by the dimension of the space in which both volumes are calculated. Or, that the Shannon entropy represents the logarithm of the effective dimension (it is going to be defined later in the text) of the distribution.

In Section II we will introduce the Born rule. The JoyStick Probability Selector is introduced in Section III. In Section IV we will present a few applications. Section V concludes the paper.

## II. QUANTUM PROBABILITY MODEL AND BORN RULE

In quantum mechanics the transition from a deterministic description to a probabilistic one is done using a simple rule termed the Born rule. This rule states that the probability of an outcome ($a$) given a state ($\Psi$) is the square of their inner product ($a^T\Psi)^2$. This section is based on a similar section in [12].

In quantum mechanics the Born rule is usually taken as one of the axioms. However, this rule has well established foundations. Gleason's theorem [1] states that the Born rule is the only consistent probability distribution for a Hilbert space structure. Wooters [13] has shown that by using the Born rule as a probability rule, the natural Euclidean metrics on a Hilbert space coincides with a natural notion of a statistical distance. Short review for some other justifications of the Born rule can be seen in [12].

The quantum probability model takes place in a Hilbert space H of finite or infinite dimension. A state is represented by a positive semidefinite linear mapping (a matrix $\rho$) from this space to itself, with a trace of 1, i.e. $\forall \Psi \in H$  $\Psi^T \rho \Psi \geq 0$, Tr($\rho$) =1.   Such $\rho$ is self adjoint and is called a density matrix.

Since $\rho$ is self adjoint, its eigenvectors $\Phi_i$ are orthonormal and since it is positive semidefinite its eigenvalues $p_i$ are real and nonegative $p_i \geq 0$. The trace of a matrix is equal to the sum of its eigenvalues, therefore $\sum_i p_i =1$.

The equality $\rho = \sum_i p_i \Phi_i \Phi_i^T$ is interpreted as "the system is in state $\Phi_i$ with probability $p_i$". The state $\rho$ is called the pure state if $\exists i$ s.t. $p_i =1$. In this case, $\rho = \Psi \Psi^T$ for some normalized state vector $\Psi$, and the system is said to be in state $\Psi$. Note that the representation of a mixed (not pure) state as a mixture of states with probabilities, is not unique if the vectors $\Phi_i$ are not mutually orthonormal.

A measurement M with an outcome $z$ in some set $Z$ is represented by a collection of positive definite matrices $\{m_z\}_{z \in Z}$ such that $\sum_{z \in Z} m_z = \mathbf{1}$ ($\mathbf{1}$ is being the identity matrix in H). Applying measurement M to state $\rho$ produces outcome $x$ with probability

$$p_z(\boldsymbol{\rho})=\text{trace}(\boldsymbol{\rho}\boldsymbol{m}_z) \ . \tag{1}$$

This is the Born rule. Most quantum models deal with a more restrictive type of measurement called the von Neumann measurement, which involves a set of projection operators $\boldsymbol{m}_a=\boldsymbol{a}\boldsymbol{a}^\text{T}$ for which $\boldsymbol{a}^\text{T}\boldsymbol{a}'=\delta_{aa'}$. In a modern language, von Neumann's measurement is a conditional expectation onto a maximal Abelian subalgebra of the algebra of all bounded operators acting on the given Hilbert space. As before, $\sum_{\boldsymbol{a}\in\text{M}} \boldsymbol{a}\boldsymbol{a}^\text{T} = 1$. For this type of measurement the Born rule takes a simpler form: $p_a(\boldsymbol{\rho})=\boldsymbol{a}^\text{T}\boldsymbol{\rho}\boldsymbol{a}$. Assuming $\boldsymbol{\rho}$ is a pure state this can be simplified further to

$$p_a(\boldsymbol{\rho}) = (\boldsymbol{a}^\text{T}\boldsymbol{\Psi})^2. \tag{2}$$

So, we can see that the probability of the outcome of the measurement will be $\boldsymbol{a}$, if the state is $\boldsymbol{\rho}$, is actually cosine square of the angle between vectors $\boldsymbol{a}$ and $\boldsymbol{\Psi}$, or $p_a(\boldsymbol{\rho})=\cos^2(\boldsymbol{a},\boldsymbol{\Psi})$.

### III.    JOYSTICK PROBABILITY SELECTOR

In this section we will give a novel but simple interpretation of the probability that is related to the Born rule. Here we will assume that we are dealing with finite dimensional discrete variable.

For the moment, let's assume that we are dealing with discrete two dimensional variables. It can associate us with a coin tossing. Assume further that two possible outcomes of our experiment are represented by the dummy variables {01} and {10}. If we represent our coin as a unit norm vector in the two dimensional space (we will call that vector JoyStick Probability Selector or JSPS), then we can have the following simple geometric interpretation given in Fig. 1.

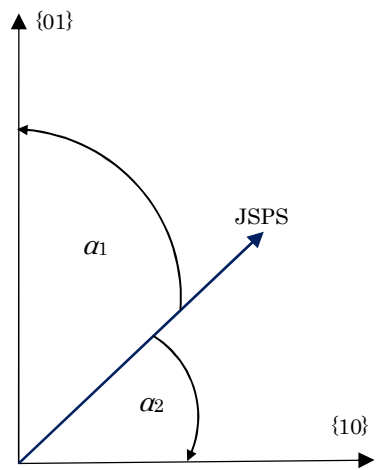

Fig 1. JoyStick Probability Selector – a 2D example

Now, we will suggest that the probability of outcome {01} is equal to cosine square of angle $α_1$, while the probability of outcome {10} is equal to cosine square of angle $α_2$. It is not difficult to see that $\cos^2(α_1) + \cos^2(α_2) = 1$. We can see that the probability of the particular outcome of the experiment (in this case coin toss) is equal to the square of the inner product of the unit norm JSPS and the unit norm vector which represents that outcome. Then, it is easy to see that this coincides with the Born rule interpretation for the case of a pure state and von Neumann measurement system.

We can check what will happen if our discrete variable is of the dimension 3. In that case our system can be represented by Fig. 2. Now, we have three possible outcomes of the experiment that are represented by dummy variables {001}, {010} and {100}. Again we have the JSPS vector which represents the status of our variable before we perform the measurement. Again, the probability of the outcome is given by the cosine square of the angle between JSPS and the particular outcome vector. It is not difficult to check that

$$\sum_{i=1}^{3} \cos^2(\alpha_i) = 1. \qquad (3)$$

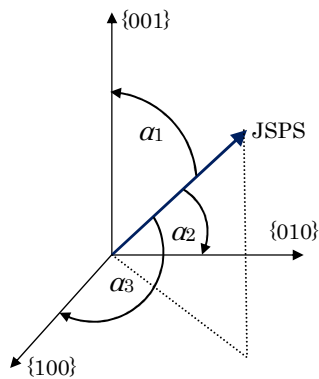

Fig. 2 A 3-D example

It follows from generalized Pythagorean Theorem, or Parseval's Theorem. For any 3-D vector whose norm is $r$ we have

$$\sum_{i=1}^{3} r^2 \cos^2(\alpha_i) = r^2. \qquad (4)$$

This way of reasoning can be extended to any finite dimension $D$. It can be extended, under some assumptions, to infinite dimensional cases (see e.g. [6] and [7]) but it will not be discussed here in detail. Here, we can see that simplex can be understood as a vector in Hilbert space of a proper dimension.

We assumed that we are working with $N$-dimensional discrete variable of finite dimension associated with a probability distribution $P=\{p_i\}$ ($i = 1,…,N$). Now, we will also assume a finite length digital representation of probabilities, and it is not difficult to show that, in that case, the probabilities can be represented by the ratio of positive integers. That means, we will assume that all $p_i$ are strictly positive – in other words our system is of a minimal possible dimension in observation (measurement) space. In that case we have

$$p_i = \frac{n_i}{d_i}, \qquad (5)$$

where $n_i$ and $d_i$ are natural numbers. If we find the least common multiple (LCM) for all $d_i$ and mark it as $D$, the probabilities can be expressed as

$$p_i = \frac{N_i}{D}, \qquad (6)$$

where

$$N_i = \frac{n_i}{d_i} D. \qquad (7)$$

It is not difficult to conclude that since $\sum_i p_i = 1$, it must be $\sum_i N_i = D$. Now, we define $D$ as intrinsic dimension of the generic space (which can be interpreted as the minimum number of experiments necessary to obtain the given discrete probability distribution), and we can associate it with a uniform distribution $P_U = 1/D$. In the generic space we have $D$ possible outcomes while the number of possible outcomes in the observation (measurement) space is $N$. In the $N$-dimensional measurement space, we have elementary events $EN_1$, $EN_2$, …, $EN_N$, while in $D$-dimensional space we have elementary events $ED_1^1$, $ED_1^2$, …, $ED_1^{N_1}$, $ED_2^1$, $ED_2^2$, …, $ED_2^{N_2}$, …, $ED_N^1$, $ED_N^2$, …, $ED_N^{N_N}$. In $D$-dimensional space any of the $N_i$ outcomes from the same group are considered different (analogy to the same particles with different spins, for instance – in measurement space we do not measure the spin, so we cannot distinguish those particles). Now, we can see the distribution $P$ as a deformed uniform distribution $P_U$ obtained by deformation of the uniform distribution in $D$-dimensional space, where deformation is based on collapsing some subspaces on their line of symmetry – the size of the subspaces is defined by numbers $N_i$, where each $N_i$ represents the number of outcomes that cannot be seen as different from the point of view of other subspaces. This will be better explained on a concrete example.

*Example*

Let's have a bent coin, such that the probability of head is two times higher than the probability of tail. In that case we can represent our system by Fig. 3. In that case, we have $\cos^2(\alpha_2)=2*\cos^2(\alpha_1)$. In that case the $p(X=H)=2/3$ and $p(X=T)=1/3$. It is easy to conclude that in this case the dimension of our generic space is 3. We can see on Fig. 3b that our two-dimensional space is obtained from 3-dimensional generic space by symmetric contraction (collapse) of the YH-BH plane onto a line defined by YH=BH. That means that we can say that we have (in the generic space) three different elementary outcomes like GreeTail, YellowHead and BlueHead. From the point of view of the tail we can see, only, the green (mixture of blue and yellow) head. So, our observational ("deformed") space is 2-dimensional because in that space we can recognize only two elementary events.

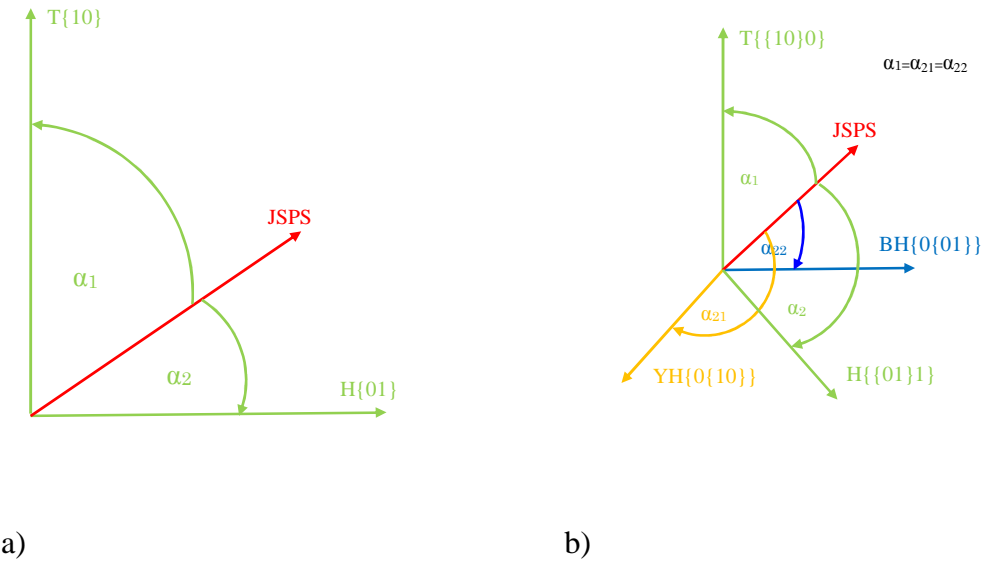

a)  b)

Fig. 3 Representation of unfair coin experiment based on JSPS

It would be difficult to visualize the proposed way of reasoning in dimensions higher than 3, but it is not difficult to conclude that a "deformed" distribution (a distribution different from uniform) can be obtained from the uniform distribution in a higher-dimensional space where some (or all) of the subspaces are collapsed to the line in the direction of diagonal of a hypercube defined by the axes of that subspace.

## IV. APPLICATIONS

The successful application of Shannon's information quantities in information theory and coding theory has stimulated the investigation of more information measures. A significant effort was made in order to make applications of information theory in other fields, as well as, to further generalize Shannon's information measure. Here we will use the proposed simple geometrical representation of probability distribution to prove some information equations in elementary manner. Also, we will show a simple way to create a code with optimal average length, just using proposed geometrical interpretation of probability distribution.

A.     *Interpretation of the Shannon entropy based on joystick probability model*

Let's assume that we are dealing with *N*-dimensional discrete probability distribution *P*={$p_1, p_2, …, p_N$}, with $0 \leq p_i \leq 1$, and $\Sigma p_i = 1$, where $p_i$ represents the probability of the system to be in the *i*-th microstate. Here we will list the three most common of the entropy measures which have found utility in a wide range of problems:

Shannon entropy [10] (also known as the Boltzmann-Gibbs entropy [8])

$$H_S(P) = -\sum_{i=1}^{N} p_i \log p_i ,$$

Rényi entropy [9]

$$H(P,r) = (1-r)^{-1} \log\left(\sum_{i=1}^{N} p_i^r\right), \quad r > 0, \quad r \neq 1,$$

and Tsallis entropy [11]

$$H(P,q) = (q-1)^{-1}\left(1 - \sum_{i=1}^{N} p_i^q\right), \quad q > 0, \quad q \neq 1.$$

For other definitions and generalization of the entropy, see e.g. [4] or [8]. The Shannon entropy is successfully applied to modern information theory, and is frequently used in other areas, such as economics, geophysics, biology, medical diagnosis and astronomy. The Rényi entropy is frequently used in many areas like coding theory and cryptography. The Tsallis entropy is used as an entropy measure in the systems in which we have presence of long-range interactions, high spatio-temporal complexity, fractal dynamics and so on. Recently, all three definitions were used in optimization problem related to Principal Component Analysis (PCA) computations [5]. In [5] it was shown that PCA could be understood, in probabilistic framework, as a constrained entropy minimization problem. It was shown that the optimal choice of entropy function, from the point of view of some optimization characteristics (e.g. convergence speed, preciseness, etc.), depends on the data that was analyzed. The Rényi and the Tsallis entropy have their geometrical interpretations related to the fractal/multifractal phase space and will not be further analyzed in this paper.

Generally speaking, any entropy function H should be (preferably) concave, invariant under permutation of $p_i$ which belongs to simplex $S_N$ of order $N$ and have maximum at the centroid of $S_N$ (i.e. when $p_i = 1/N$ for all $i$), and minima at the vertices of $S_N$ (i.e. when $p_i$ are maximal different – for instance when one $p_i$ is almost 1, and all others are as small as possible). This gives us a lot of possibilities for the entropy definition that could be used in optimization problems. Which definition is used, should depend on the target application. In the case we want to create a new entropy measure for the interpretation of some physics processes we should have in mind some restrictions that are related to the specific process. For instance, although the function $H_S(P)^5$ is quite correct form the point of view of the optimization process, it is not clear that it has any relevance in physics, and it is not likely that physicists are going to use it. In this section we are going to give an interpretation of the Shannon entropy from the "geometrical" point of view. Here, we rely on some definitions given in the previous section and we will add some new ones. We assume the following:

- $N$ is the number of elementary outcomes of the experiment in the measurement space.

- $D$ is the generic dimension of the distribution. We will also define $D$ as the length of the absolutely typical ensemble (this definition will be given in the next paragraph). We define the generic uniform distribution as a uniform distribution with $D$ possible elementary outcomes (we consider that the same $N_i$ elementary outcomes $EN_i$ in the measurement space can be seen as $N_i$ different $ED_i^{N_i}$ outcomes in the generic space).

- Absolutely typical ensemble (ATE) is a sequence of length $D$ which consists of exactly $N_i$ elementary outcomes from each of the $ED_i^{N_i}$ group.

- The combinatorial volume of the typical ensemble in the measurement space is defined as

$$V_{\text{info}} = \prod_{n=1}^{N} N_n^{N_n}. \tag{8}$$

- The combinatorial volume of the generic uniform distribution is expressed as

$$V_{\text{uinfo}} = D^D. \tag{9}$$

- The ratio of the combinatorial volume of the generic uniform distribution and the combinatorial volume of the absolutely typical ensemble is (see (6) and (7) and text bellow them)

$$R = \frac{V_{\text{uinfo}}}{V_{\text{info}}} = \frac{D^D}{\prod_{n=1}^{N} N_n^{N_n}} = \prod_{n=1}^{N}\left(\frac{D}{N_n}\right)^{N_n} = \left(\prod_{n=1}^{N}\left(\frac{1}{\frac{N_n}{D}}\right)^{\frac{N_n}{D}}\right)^D$$

$$= b^{\log_b\left(\left(\prod_{n=1}^{N}\left(\frac{1}{\frac{N_n}{D}}\right)^{\frac{N_n}{D}}\right)^D\right)} = b^{D\log_b\left(\left(\prod_{n=1}^{N}\left(\frac{1}{p_n}\right)^{p_n}\right)\right)}$$

$$= b^{-D\sum_{n=1}^{N} p_n \log_b(p_n)} = b^{D\mathrm{H}(P)}, \tag{10}$$

where H($P$) some entropy function of discrete distribution that is analyzed and $b$ is a positive integer that represents a base for the logarithm. The Shannon entropy $H_S$, represents special case when we have $b=2$. So, we can see that the Shannon entropy can be interpreted as

$$\mathrm{H_S}(P) = \frac{1}{D}\log_2(R) = \log_2\left(R^{1/D}\right) = \log_2(eff\_dim\,(P)), \tag{11}$$

where $R$ represents the ratio of the combinatorial volume of the generic uniform distribution (GUD) related to given discrete distribution and the combinatorial volume of absolutely typical ensemble drawn from the given distribution. In other words, the entropy is inversely proportional to the volume of the hypercuboid defined by the absolutely typical ensemble. The other interpretation is that it is proportional to the number of hypercuboids defined by the ATE that can be stored in the hypercube defined by the GUD. And, since we can see that $R^{1/D}$ is defined as the *effective dimension* of the distribution *eff_dim*(*P*), the Shannon entropy can be seen as a logarithm of the effective dimension of a distribution. In order to understand the meaning of the effective dimension, we are going to calculate it for several examples of distributions. We will use 2-dimensional observational distribution space and we will calculate the effective and generic dimensions. The results are given in the following table:

Table I

| Distribution | Generic dimension | Effective dimension |
| --- | --- | --- |
| P(1/2, 1/2) | 2 | 2 |
| P(1/4, 3/4) | 4 | 1.7548 |
| P(1/16, 15/16) | 16 | 1.2634 |
| P(1/256, 255/256) | 256 | 1.0259 |

We can clearly see that easily predictive outcomes have lower effective dimension which tends toward one. If the effective dimension is one, we can talk about absolutely predictive outcome.

Essentially, the entropy represents the ratio of the combinatorial volume of the generic uniform distribution and the combinatorial volume of absolutely typical ensemble. If we want to interpret it as an average number of bits required for coding the symbols, and later decoding them uniquely after transmission trough the noiseless channel, we should take logarithm of the ratio *R* and divided it by the dimension of the generic space, which coincides with the definition of Shannon's entropy. Of course, all this can be extended to the Boltzmann-Gibbs entropy. In the case of the Boltzmann-Gibbs entropy, it is clear that the dimension of the generic space cannot be larger than the number of particles.

Here we will point out that usual introduction of the entropies in the machine learning books (see e.g. [2] page 51) are slightly misleading, since they try to explain everything from the point of view of the measurement space, although it (probably) should be done in the generic space.

*Note*: In the proposed context, it seems, it is difficult to give geometrical interpretation of Kullback-Leibler divergence.

*Simple example*

Now, we are going to illustrate proposed geometrical interpretation of entropy on a very simple example. Again, let's have a bent coin, such that the probability of head is two times higher than the probability of tail. If we have in mind representation given in Fig. 3, we can easily concluded that $\cos^2(\alpha_2)=2*\cos^2(\alpha_1)$. In that case the $p(X=H)=2/3$ and $p(X=T)=1/3$. That means that in this case the length of ATE is 3. We can see on Fig. 3b that our two-dimensional space is obtained from the 3-dimensional generic space by symmetric contraction of the YH-BH plane into a line defined by YH=BH. That means that we can say that we have (in the generic space) three different elementary outcomes like GreeTail, YellowHead and BlueHead. From the point of view of the tail we only see the green head (so our "deformed" space is 2-dimensional because in that space we can recognize only two elementary events). So, our ATE can be represented by all possible combinations of three elementary events, where two of them are heads and one is tail. The information volume of the ATE is $V_{info}=2^2 1^1$, so it can be visualized as cuboid of dimensions 2, 2 and 1. Obviously it represents a number of all possible variations with repetitions in subspaces of dimensions 2 and 1. Our GUD is of dimension 3 and its information volume is $V_{uinfo}=3^3$. We can visualize it as a cube of dimension 3. So, geometrical representation of the information volumes in the concrete case can be depicted as it is done in Fig 4. We can see that the entropy is proportional to the number of cuboids that can be put in the cube. This gives very simple geometrical interpretation of the Shannon noiseless coding theorem. Also, it will be simple to give simple geometrical interpretation of the conditional entropy in the case of two variables. In that case the conditional entropy would be proportional to the ratio of the ATE volume of one variable that is not shared with another variable and volume of GUD. In the same manner, the mutual information could be interpreted as proportional to the ratio of the ATE volume shared by variables and volume of GUD.

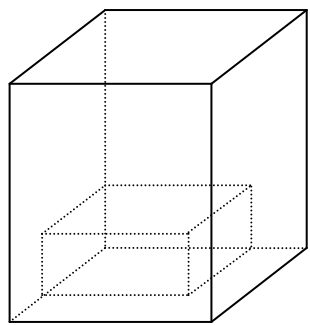

Fig. 4 – Geometrical interpretation of information volumes defined by ATE (cubiod) and GUD (cube)

Here we will make elementary proofs for two nonequalities that are frequently used in information theory.

A) It is very simple to show that H(X) ≥ H(X|Y) – it is a simple geometrical property that says that the volume defined by variable X can be contained partially or completely by the volume defined by the variable Y.

B) From A), it immediately follows that mutual information is nonnegative, or I(X, Y) ≥0. Also, it is not difficult to see that I(X, Y) = I(Y, X) since, the volume defined by X that is contained in the volume defined by Y is equal to the volume defined by Y contained in the volume defined by X. Furthermore, if variable X and Y are independent they do not share any volume, and consequently the mutual information becomes zero.

*B.    Optimal coding length*

Here, we will show how the proposed geometrical interpretation can be used for the purpose of coding of an input sequence with optimal code length. As an example we will use the example used in seminal Shannon's paper [10]. Consider the system which has 4 possible elementary outcomes A, B, C and D with probabilities 1/2, 1/4, 1/8, and 1/8, respectively.  It is not difficult to calculate that Shannon's entropy is $H_S$=7/4. So, we need 7/4 bits per symbol, on average, if we want to code the message. How can we achieve this code? We can notice that the dimension of our system is $N$=4. Also, from the value of the probabilities $p_i$ we can easily conclude that dimension of the generic space (system) is $D=8$. Also, we can see that N1=4, N2 =2, N3 =1 and N4 = 1, which means that in the generic space we have 8 symbols A1, A2, A3, A4, B1, B2, C and D. In the original space and from the point of view of other symbols, symbols A1, A2, A3 and A4, as well as, B1 and B2 are indistinguishable. So, let's make, initially, the usual code for 8 uniformly distributed symbols in a generic space. In that case we can have the following

A1 - 000

A2 - 001

A3 - 010

A4 - 011

B1 - 100

B2 - 101

C - 110

D - 111.

Since A1, A2, A3 and A4 are not distinguishable from the point of view of other symbols, we can conclude that the variable part of those symbols code is irrelevant, and the only constant part in all Ai symbols code is leading 0. So, we can code A as 0. Using the same way of reasoning, we can see that B should be coded with 10. The symbols C and D should be coded as 110 and 111, respectively. So, we can see that the average code length is

1/2*1+1/4*2+1/8*3+1/8*3 = 7/4,

and it is the length that can be calculated by Shannon's entropy.

*C.     Projection entropy – a new entropy measure*

Lets define the following ratio $R_{\text{PInfo}}$ in the *N*-dimensional space as

$$R_{\text{PInfo}} = \prod_{i=1}^{N} p_i = \frac{\prod_{i=1}^{N} N_i}{D^N}. \qquad (12)$$

We can see that it represents the ratio of the volume of the hypercuboid with edges lengths $N_i$ and the volume of the hypercube of edge length *D* in *N*-dimensional space. If we fix the number of elementary outcomes in the observational space, we can conclude that value $\sqrt[N]{R_{\text{PInfo}}}$ represents a relative entropy function – it can be used as a relative entropy measure for different distribution of the same dimension, but it cannot be used as a global entropy measure. If we want to make a global entropy measure, we can define the projection entropy as

$$H_P(P) = \log\left(N^2 \sqrt[N]{R_{\text{PInfo}}}\right). \tag{13}$$

From (13) we can easily conclude that the function $H_P$ has several properties that are very similar to the Shannon entropy:

1. $H_P$ is continuous in the $p_i$.

2. If all $p_i$ are equal, $p_i=1/N$, then $H_P$ is a monotonically increasing function of $N$. With equally likely events there is more choice, or uncertainty, when there are more possible events.

3. It is not difficult to check that if $P$ and $Q$ represents two independent distributions then $H_P(P, Q) = H_P(P) + H_P(Q)$.

4. The Shannon entropy has the property that if the choice is broken into two successive choices, the original H should be the weighted sum of the individual values of H. This can be formulated as exact rule by the following equation:

$$H(p_1, \ldots, p_{n-1}, q_1, q_2) = H(p_1, \ldots, p_{n-1}, p_n) + p_n H\left(\frac{q_1}{p_n}, \frac{q_2}{p_n}\right)$$

where $p_n = q_1 + q_2$. In the case of the projection entropy that rule can be expressed as

$$H_P(p_1, \ldots, p_{n-1}, q_1, q_2)$$

$$= \frac{n}{n+1} H_P(p_1, \ldots, p_{n-1}, p_n) + \frac{2}{n+1} H_P\left(\frac{q_1}{p_n}, \frac{q_2}{p_n}\right) + \frac{1}{n+1} \log p_n$$

$$+ 2\log(n+1) - \frac{n}{n+1} 2\log(n) - \frac{2}{n+1} 2\log 2,$$

where, again, $p_n = q_1 + q_2$.

The projection entropy $H_P$ has some features that are different from the Shannon entropy, like:

- It can take negative values. This will happen in the cases when one or several outcomes have very small probabilities.

- The value $N^2 \sqrt[N]{R_{PInfo}}$ cannot be interpreted as an effective dimension of a distribution.

- Number of calculations for the projection entropy is smaller than the number of calculations for the Shannon entropy. This could be useful in some optimization applications.

*Note*: It is not difficult to notice that $\log(R_{PInfo})$ represents a log-likelihood function.

## V.  CONCLUSION

In this paper we proposed a geometrical interpretation of the discrete probability based on, recently proposed JSPS model. We were exclusively dealing with discrete probability distributions of finite dimension which can be represented by finite length digital representation. Based on JSPS model, it was shown that the Shannon entropy essentially represents the ratio of the combinatorial volume of the generic uniform distribution (GUD) and the combinatorial volume of absolutely typical ensemble (ATE). It was shown how some inequalities form information theory could be proved in elementary manner. We also proposed a new entropy measure that is calculated in the measurement (observational) space.